\documentclass[traditabstract]{aa}
\usepackage{graphicx}
\usepackage{txfonts}
\usepackage{epsfig}
\usepackage{natbib}
\usepackage{color}
\begin{document}

%
\title{Radial migration in galactic disks caused by resonance overlap of multiple
patterns: Self-consistent simulations}

 \author{I.~Minchev\inst{1,2}, B.~Famaey\inst{2,3}, F.~Combes\inst{4}, P.~Di~Matteo\inst{5}, M.~Mouhcine\inst{2,6}, H.~Wozniak\inst{2}
}
\institute{AIP, An der Sterwarte 16, 14482 Potsdam, Germany
\and
Observatoire Astronomique de Strasbourg, CNRS UMR 7550, 
67000 Strasbourg, France
\email{ivan.minchev@astro.unistra.fr}
\and
AIfA, Universt\"at Bonn, 53121 Bonn, Germany
\and
Observatoire de Paris, LERMA, 61 avenue de L'Observatoire,75014 Paris, France
\and
Observatoire de Paris-Meudon, GEPI, CNRS UMR 8111, 5 pl. Jules Janssen, Meudon, 92195, France
\and
ARI, Liverpool John Moores University, Twelve
Quays House, Egerton Wharf, Birkenhead, CH41~1LD, UK}


\abstract{
We have recently identified a new radial migration mechanism resulting from the
overlap of spiral and bar resonances in galactic disks. Here we confirm the
efficiency of this mechanism in fully self-consistent, Tree-SPH simulations, as
well as high-resolution pure N-body simulations. In all barred cases we clearly
identify the effect of spiral-bar resonance overlap by measuring a bimodality in 
the changes of angular momentum in the disk, $\Delta L$, whose maxima are near the 
bar's corotation and outer Lindblad resonance. This contrasts with the smooth
distribution of $\Delta L$ for a simulation with no stable bar present, where
strong radial migration is induced by multiple spirals. The presence of a disk
gaseous component appears to increase the rate of angular momentum exchange by 
about 20\%. The efficiency of this mechanism is such that galactic stellar disks 
can extend to over 10 scale-lengths within 1-3~Gyr in both Milky Way size and 
low-mass galaxies (circular velocity $\sim100$~km/s). We also show that metallicity
gradients can flatten in less than 1~Gyr rendering mixing in barred galaxies an 
order of magnitude more efficient than previously thought.
}

\titlerunning{Mixing in galactic disks caused by multiple patterns}
\authorrunning{I. Minchev et al.}

\maketitle

\section{Introduction}
\label{sec:intro}

In the past few decades, discrepancies in the solar neighborhood age-metallicity relation have 
implied that effective radial migration (i.e., redistribution of angular 
momentum) must be taking place in the Milky Way disk (\citealt{edvardson93,
haywood08, schonrich09}; see \citealt{mf10} for a comprehensive discussion). 
In parallel, there is now considerable observational evidence that the non-axisymmetry of 
the Galactic potential can cause significant perturbations in the motion of stars of all 
ages. Evidence for this comes from, e.g, the moving groups in the solar neighbourhood 
\citep{dehnen98} containing stars of very different ages \citep{famaey05} or the non-zero 
value of the $C$ and $K$ Oort constants for red giant stars in the extended local disk 
\citep{od03, siebert10}. All of these can be explained by the effect of resonances associated
with a central bar \citep{minchev10,mnq07} or spiral structure (SS) \citep{qm05,quillen10}.

Until recently it was accepted that efficient radial mixing of stars in galactic disks 
was caused solely by transient spirals \citep[][hereafter SB02]{sellwood02}. However, 
\cite{quillen09} showed that 
small satellites on radial, in-plane orbits can cause mixing in the outer disk and thus 
account for the fraction of low-metallicity stars present in the solar neighborhood 
\citep{haywood08}. Moreover, we have recently demonstrated (\citealt{mf10}, hereafter MF10) that 
a strong exchange of angular momentum occurs when a stellar disk is perturbed by a central 
bar and SS simultaneously: our test-particle simulations allowed us to 
attribute this effect to the overlap or first and second order resonances of each perturber. 
The presence of multiple patterns in galactic disks has been observed both in external 
galaxies (e.g., \citealt{elmegreen92}) and N-body simulations (e.g., 
\citealt{rautiainen99}). Given that more than two-thirds of disk galaxies, including our own 
Milky Way, contain both central bars and SS, it is imperative to establish a strong 
understanding of the implications of this mechanism. 
Therefore, in this work we study a range of fully self-consistent, Tree-SPH simulations, 
as well as high resolution pure N-body simulations, searching for the signature of spiral-bar
interaction in the distribution of angular momentum: MF10 predicted that when both bar and 
spirals are present in the disk, one finds that the changes in angular momentum form a bimodal 
distribution, with two local maxima close to the corotation and outer Lindblad resonance 
of the bar, regardless of the pattern speed of the spiral.

\label{sec:galmer}
\begin{figure*}
\includegraphics[width=17cm]{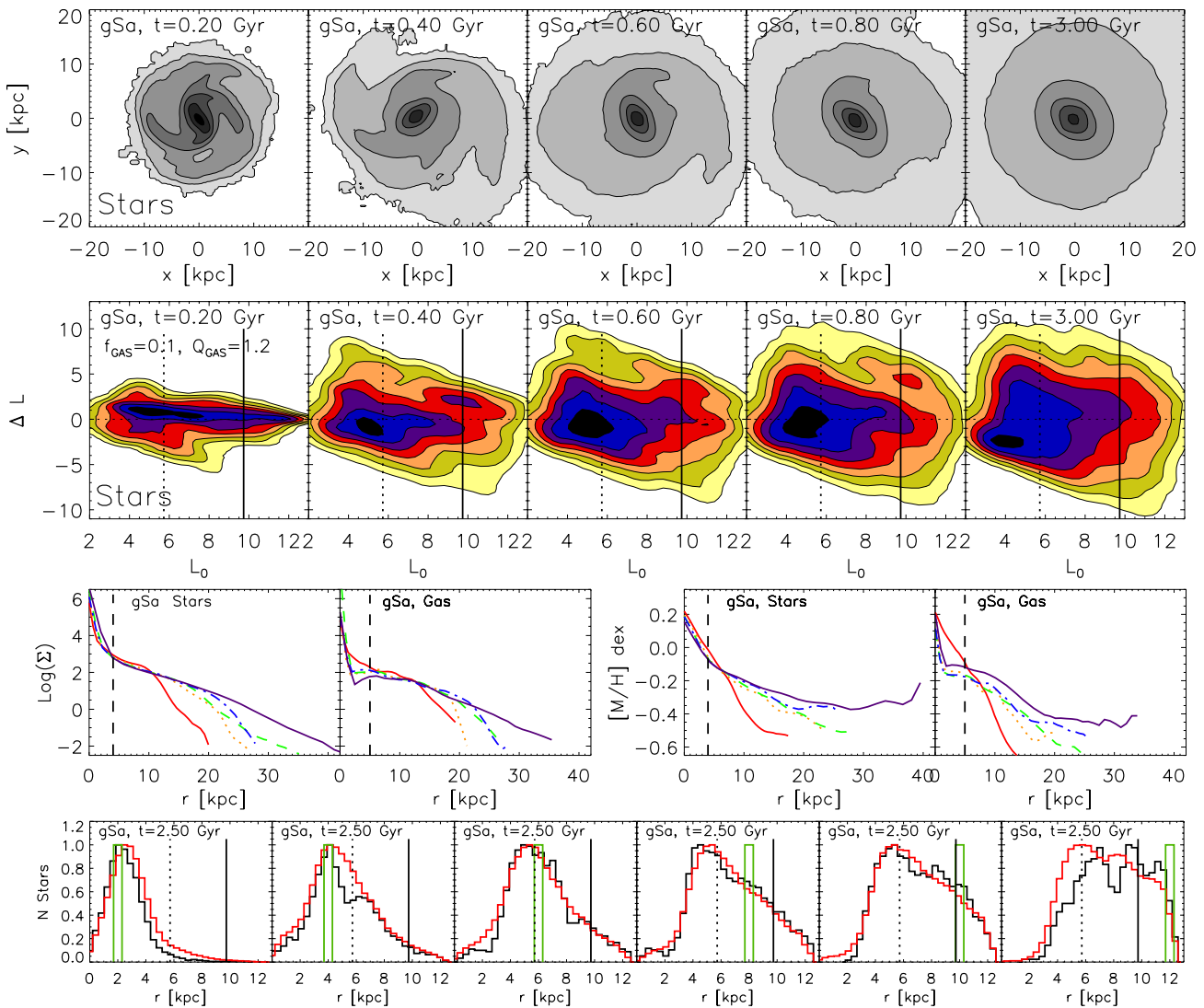}
\caption{
First row: Stellar disk number density contours of the isolated giant Sa galaxy simulation
in GalMer for 5 time outputs as indicated in each panel. Second row: Changes in angular 
momentum, $\Delta L$ as a function of the initial angular momentum, $L$. Both $\Delta L$
and $L$ are divided by the asymptotic circular velocity $V_c$, thus $L$ gives approximately 
the galactic radius in kpc. The locations of the bar's corotation and 2:1~OLR are indicated 
by the dotted and solid lines, respectively. A bimodal distribution indicating
the work of resonance overlap of bar and spirals is clearly seen in each panel. 
Third row: The evolution of the radial profiles (left) and metallicities (right) for the 
stellar and gaseous disks. The initial stellar and gaseous disk scale-lengths are 
indicated by the dashed lines. The time steps shown are as in the first row, indicated by
solid red, dotted orange, dashed green, dotted-dash blue, and solid purple lines, respectively.
Fourth row: Distribution of birth radii for stars ending up in the annuli indicated by the
green lines (600 pc) at t=2.5~Gyr. 
 }
\label{fig:gSa}
\end{figure*}

\section{Models and results}

While not self-consistent, test-particle simulations allow for a full control over
the simulation parameters, such as the amplitudes and pattern speeds of bar and SS,
and still provide a good approximation to self-consistent simulations. Employing 
this method in MF10, we were able to suppress the effect of transient spirals and thus 
identify the non-linear effect of resonance overlap. We showed that the most 
important signature of this mechanism was a bimodality in the changes in angular 
momentum in the disk, $\Delta L$, with maxima near the bar's corotation and its outer 
Lindblad resonance (OLR) regardless of the SS pattern speed. 

Hereafter, we analyze self-consistent 
simulations with strong bars and spirals, therefore mixing from {\it both} transient 
spirals (SB02) and resonance overlap (MF10) is expected. However, we hope to be 
able to identify the latter mechanism by detecting the aforementioned bimodality, given that 
the predicted distribution of $\Delta L$ generated by transient spirals without a bar 
is rather smooth and not bimodal (see SB02). 

\subsection{Tree-SPH simulation of a giant Sa spiral galaxy}

\begin{figure*}
\includegraphics[width=16.5cm]{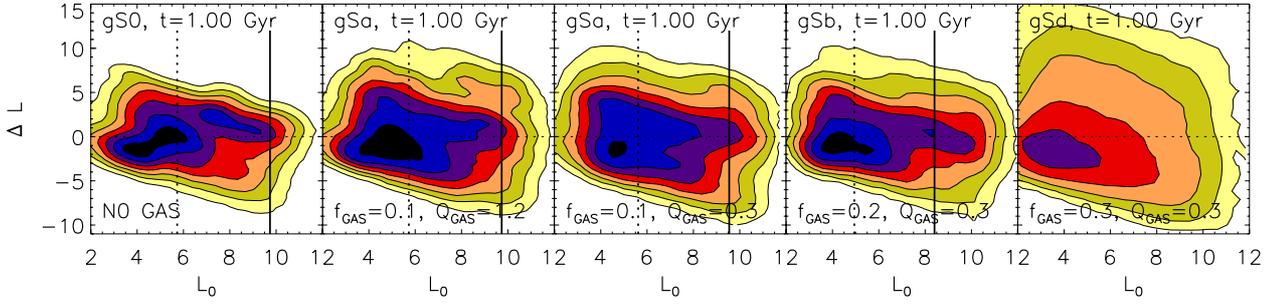}
\caption{
Same as the second row of Fig.~\ref{fig:gSa} but for all isolated giant GalMer galaxies. 
Contour levels are the same for all plots and the time is t=1~Gyr. The locations of the 
bar's corotation and OLR are indicated by the dotted and solid lines, respectively. 
The rightmost panel shows the only disk lacking a central bar. In this case, resonance overlap
from strong multiple spirals gives rise to a smooth $\Delta L$ distribution. 
 }
\label{fig:gSall}
\end{figure*}
We test the predictions of MF10 by analyzing fully self-consistent simulations of isolated 
disk galaxies from the GalMer database\footnote{http://galmer.obspm.fr}, including a gas 
component as well as star formation \citep{dimatteo07,chilingarian10}. 

In Galmer, for each galaxy type, the halo and the optional bulge are modeled as Plummer 
spheres, with characteristic masses $M_H$ and $M_B$, and characteristic radii $r_H$ and 
$r_B$, respectively. Their  densities are given by:
\begin{equation}\label{halo}
\rho_{H}(r)=\left(\frac{3M_{H}}{4\pi {r_{H}}^3}\right)\left(1+\frac{r^2}{{r_{H}}^2}\right)^{-5/2}
\end{equation}
and
\begin{equation}\label{bulge}
\rho_{B}(r)=\left(\frac{3M_{B}}{4\pi {r_{B}}^3}\right)\left(1+\frac{r^2}{{r_{B}}^2}\right)^{-5/2}.
\end{equation}
On the other hand, the gaseous and stellar disks follow Miyamoto-Nagai density profiles with masses $M_{g}$ and  $M_{*}$ and  vertical and radial scale lengths given, respectively, by $h_{g}$ and $a_{g}$, and $h_{*}$ and $a_{*}$: 
\begin{eqnarray}\label{gasdisk}
\rho_{g}(R,z)&=&\left(\frac{{h_{g}}^2 M_{g}}{4 \pi}\right)\times\nonumber\\&&\frac{a_{g} R^2+(a_{g}+3\sqrt{z^2+{h_{g}}^2})\left(a_{g}+\sqrt{z^2+{h_{g}}^2}\right)^2}
{\left[a_{g}^2+\left(a_{g}+\sqrt{z^2+{h_{g}}^2}\right)^2\right]^{5/2}\left(z^2+{h_g}^2\right)^{3/2}},
\end{eqnarray}
\begin{eqnarray}\label{stdisk}
\rho_{*}(R,z)&=&\left(\frac{{h_{*}}^2 M_{*}}{4 \pi}\right)\times\nonumber\\&&\frac{a_{*} R^2+(a_{*}+3\sqrt{z^2+{h_{*}}^2})\left(a_{*}+\sqrt{z^2+{h_{*}}^2}\right)^2}
{\left[a_{*}^2+\left(a_{*}+\sqrt{z^2+{h_{*}}^2}\right)^2\right]^{5/2}\left(z^2+{h_*}^2\right)^{3/2}}.
\end{eqnarray}

Here, we first analyze the properties of the isolated giant Sa (gSa) galaxy model: it has a total mass of $\sim 2.4\times10^{11} M_{sun}$, with $M_H=1.15\times10^{11} M_{sun}$, $M_B=0.23\times10^{11} M_{sun}$, $M_g=0.09\times10^{11} M_{sun}$, and $M_{*}=0.92\times10^{11} M_{sun}$. The scale radii are $r_H=10$~kpc, $r_B=2$~kpc, $h_{g}=0.2$~kpc, $a_{g}=5$~kpc, $h_{*}=0.5$~kpc, and $a_{*}=4$~kpc. The initial Toomre parameter of both stars and gas is taken to be $Q=1.2$ as the initial condition of the Tree-SPH simulations. Fig.~\ref{fig:gSa} shows the properties of this isolated gSa model as a function of time. 

\subsubsection{Identification of the bimodality}

In the first row of Fig.~\ref{fig:gSa}, we plot stellar number density contours for five time 
outputs up to 3 Gyr as indicated in each panel. We note that the initially strong SS disappears 
by the end of the simulation. The expansion of the disk is already an indication of radial 
migration. 

The second row of Fig.~\ref{fig:gSa} shows contours of the changes in the angular momentum of 
stars, $\Delta L$, versus the initial angular momentum, $L_0$. Both $\Delta L$ and $L$ are 
divided by the asymptotic circular velocity $V_c\approx280$~km/s, thus $L$ gives approximately 
the galactic radius in kpc. The bar's corotation and 2:1~OLR are indicated by the dotted and 
solid lines. A bimodal distribution becomes apparent as soon as the bar and SS form 
($t\approx0.2$ Gyr). While the peak at $L_0\approx4.5$ is caused by the bar's corotation, 
the one at $L_0\approx8.5$ could {\it only} be produced by resonance overlap 
(see Fig.~2 in MF10). At all later times this clear bimodality remains in the distribution 
of $\Delta L$, thus providing strong evidence that the effect we see is caused by the spiral-bar 
interaction. We note also that the mixing timescale here is simply too short to be the effect 
of transients only.

\subsubsection{Extension of the disk}

The third row of Fig.~\ref{fig:gSa} shows the temporal evolution of the radial density 
(left) and metallicity (right) profiles for both the stellar and gaseous disks. The time 
steps shown are as in the first row, indicated by solid red, dotted orange, dashed green, 
dotted-dash blue, and solid purple in increasing order. 

Owing to the vigorous mixing, the disk grows rapidly with time displaying a break in the 
density profile that moves progressively to larger radii. At the final time (purple lines), 
the stellar and gaseous disks have extended to about ten and seven scale-lengths, respectively, 
while 
roughly preserving their exponential profiles. This is accompanied by a strong flattening 
in the metallicity gradients, where at t=3~Gyr one can clearly see a reversal in
the metallicity gradient of stars in the outer region of the disk ($\sim30$~kpc).
The stellar metallicities at that location are similar to the initial ones at 8~kpc, 
suggesting that stars in the outskirts of the disk originate from regions near the bar's 
2:1~OLR, where the changes in $\Delta L$ are most prominent. 

\subsubsection{Birth radii of stars}

The bottom row of Fig.~\ref{fig:gSa} shows the distribution of birth radii for stars ending 
up in the annuli indicated by the green lines (600~pc) at $t=2.5$~Gyr. The black and red 
histograms show stars on nearly circular orbits (of velocity dispersion smaller than 10~km/s) 
and the total population, respectively. In all cases, a large fraction of stars are found to
originate from the bar's corotation (dotted line), as well as an increasingly large 
fraction from the bar's OLR (solid line) as larger final radii are sampled.  

\subsection{Tree-SPH simulations of other galaxy types}

To assess how our results depend on the peculiar galaxy type we considered, 
we plot in Fig.~\ref{fig:gSall} the changes in angular momentum for the stellar disk
components of {\it all} GalMer isolated giant galaxy models available 
\citep[Table~1]{chilingarian10}. Contour levels are the same for all plots and the time is 
t=1~Gyr. All apart from the gSd (bulgeless) model develop long-lived, central bars at the 
beginning of the simulation and consequently show a bimodality in $\Delta L$. When we compare 
the gS0 and gSa models, which have identical initial conditions except that gS0 is gas-free, 
we find that $\Delta L$ increases by $\sim$~20\% when gaseous disk component is introduced.
The stability of the gas, however, does not appear to play a major role. Only a small difference 
occurs when the gas is made more unstable when $Q_{gas}$ is lowered from 1.2 to 0.3
(second and third panels), where $Q_{gas}$ is the Toomre instability parameter for the gas. 
The weaker bar in the gSb model (fourth panel) results in reduced mixing
as we predicted in MF10. Unlike all other models, the gSd simulation (rightmost 
panel) does not form a stable bar. The strong mixing observed here results from the large 
number of multiple spirals \citep{elmegreen92,rix93,mq06} propagating simultaneously through 
the disk during the first $\sim500$~Myr. Due to the various SS pattern speeds,
there are no radii at which the effect of resonance overlap differs distinctly
from the rest of the disk (such as the bar's corotation and OLR).
Consequently, we observe a smooth distribution of $\Delta L$ in contrast to the
case of bar + SS.

\begin{figure}
\includegraphics[width=8.5cm]{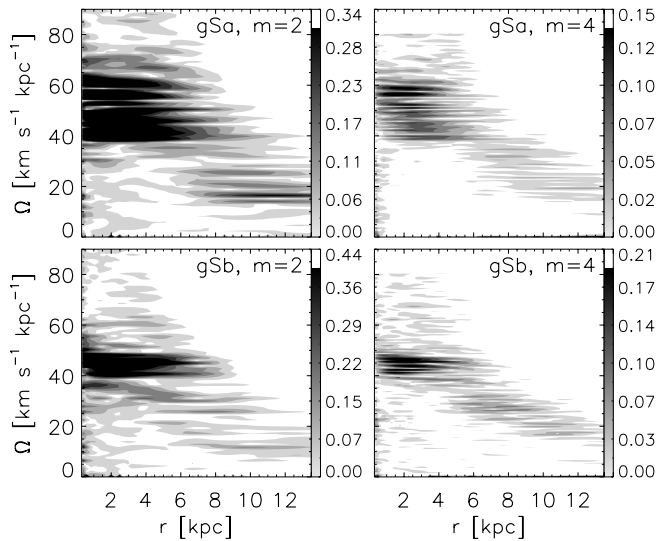}
\caption{
Power spectrum displaying the pattern speed of the $m=2$ and $m=4$ Fourier components,
during the entire simulation, of the gSa (top) and gSb (bottom) models. Contour levels are 
indicated for each panel. Note that the gSa bar evolves (slows down and extends) much 
more and its two-armed SS is twice as strong as for the gSb model (top left). Despite 
the slightly stronger gSb bar, the effect on $\Delta L$ in the gSa's outer disk is much more 
prominent (Fig.\ref{fig:gSall}) because of the combined effect of the bar and stronger spirals.  
 }
\label{fig:omega}
\end{figure}

\subsection{Pattern speeds}

Fig. \ref{fig:omega} shows a power spectrum displaying the pattern speeds of the $m=2$ and 
$m=4$ (two-armed and four-armed structure) Fourier components, over the whole simulation, of 
the gSa (top) and gSb (bottom) models. Contour levels are indicated for each panel in units 
of $Q_T$, the ratio of the maximum tangential force to the azimuthally averaged radial force 
at a given radius. For both simulations, the SS ($r\ga5$ kpc) spans a range of pattern 
speeds, always slower than the bar. As expected from the results of MF10, the SS pattern speed 
has little influence on the location of the maxima in the bimodal distribution of $\Delta L$. 
Note that the gSa bar evolves (slows down and extends over time) much more, and that its 
two-armed SS is twice stronger compared to the gSb model (top left). Despite the slightly 
stronger gSb bar, the effect on $\Delta L$ for the gSa is much more prominent in the outer 
disk (Fig.\ref{fig:gSall}) because of the combined effect of the bar and stronger spirals. 
Strong spirals are thus needed for this migration mechanism to be efficient.
However, the relation between the strength of the perturbers and the
angular momentum redistribution is nonlinear as previously shown by MF10. 

\subsection{High resolution N-body simulations}
\label{sec:nbody}

\begin{figure}
\includegraphics[width=7cm]{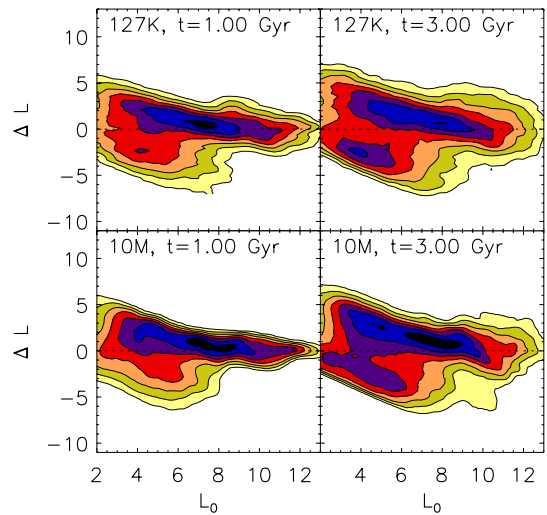}
\caption{
Contrasting the effect on $\Delta L$ for a low-resolution ($1.27\times10^5$) and a 
high-resolution $10^7$ pure N-body simulations.
 }
\label{fig:nbody}
\end{figure}

To determine whether the strong mixing we described above is caused by a deficiency in the 
resolution of the GalMer simulations, we ran pure N-body collisionless simulations with $10^7$
and $1.27\times10^5$ particles in the disk. Full description and details can be found in
\cite{wozniak09}. With a scale-length of 3.5 kpc, the initial conditions of the disk 
are comparable to those of the gSb GalMer model (Fig.~\ref{fig:gSall}) but 
lack a gas component. These simulations have no halo, which causes a $\sim30\%$ 
drop in the RC at $\sim10$~kpc. 
 
Fig.~\ref{fig:nbody} shows the changes in angular momentum at t=1 and 3~Gyr for 
$1.27\times10^5$ (top) and $10^7$ (bottom) disk particles. At the beginning of the 
simulations, the effect is stronger for the low-resolution case due to the faster bar 
formation. However, at $t\approx2$~Gyr both the high- and low-resolution runs yield
a similar result. This suggests that the magnitude of the radial migration induced
by the bar-spiral interaction and the timescale over which it is effective are not strongly
dependent on the numerical resolution of the simulations. We conclude that the properties
of the mixing, i.e., its magnitude and timescale, that we infer from the GalMer simulations
are only weakly affected by the insufficient resolution.

\subsection{Migration in low-mass galaxies}

\begin{figure}
\includegraphics[width=8.7cm]{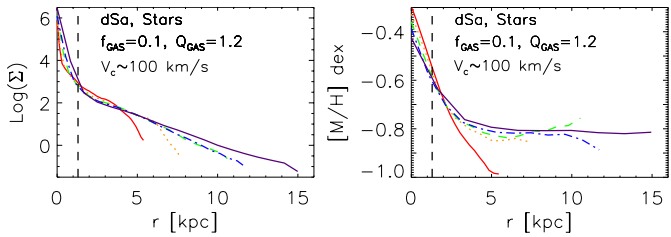}
\caption{
The time evolution of the stellar disk density (left) and metallicity (right) profiles for 
the dwarf Sa isolated GalMer model. Time outputs and colors are the same as the third row of 
Fig.~\ref{fig:gSa}. This is a low-mass galaxy with a rotation curve of $\sim100$~km/s 
similar to NGC~300 and M33. The initial disk scale-length is indicated by the dashed line.
The disk extends to over ten scale-lengths in less than 3~Gyr. 
 }
\label{fig:dSa}
\end{figure}

We now investigate whether the resonance overlap mechanism is also efficient in low-mass 
galaxies. \citet{gogarten10} showed that although transient spirals can explain extended 
disks for MW-type galaxies, this mixing mechanism is inefficient for galaxies with RCs of
$V_c\sim100$~km/s, such as NGC~300 and M33. Nevertheless, both NGC~300 and M33
are observed to have extended radial profiles of up to ten scale-lengths. We now
consider the barred, dwarf Sa (dSa) simulation in GalMer (RC of 100~km/s and initial 
scale-length of 1.3 kpc) to see how its density distribution and metallicity gradient 
evolve with time. In Fig.~\ref{fig:dSa} we plot the time development of the stellar disk 
density and metallicity profiles, for the same time steps as in Fig.~\ref{fig:gSa}.
We can clearly see that the stellar disk extends to more than ten scale-lengths in 
$\sim3$~Gyr, while preserving its exponential mass density profile. At the same time, the
metallicity gradient becomes flat in less than 1~Gyr. 
Both the predicted extent of the stellar disk and the flattening of the metallicity profile
are consistent with what has been recently reported for nearby low-mass galaxies
\citep{bland05,vlajic09}.

\section{Conclusions}

We have examined the redistribution of angular momentum in galactic disks by means of 
Tree-SPH and high-resolution pure N-body simulations. We have found that resonance overlap 
of multiple patterns (such as bar + SS or SS + SS) induces strong exchange of angular momentum 
throughout the disks in agreement with the predictions of MF10. Since in the self-consistent 
simulations 
analyzed in this work spirals may be transient, we should also expect a contribution from the 
SB02 radial migration mechanism. However, for the short timescales considered here 
($<1$~Gyr) transients would simply have a very small effect (SB02). The resonance overlap 
mechanism is clearly identified by a bimodality in the changes in angular momentum, 
$\Delta L$, caused by the bar's corotation and 2:1~OLR (Fig.~\ref{fig:gSa}, \ref{fig:gSall}, 
and \ref{fig:nbody}). We have contrasted this to a simulation lacking a stable central bar 
(Fig.~\ref{fig:gSall}, rightmost panel), where the $\Delta L$ distribution is smooth. 
The effect is especially strong when a gaseous component is present, as a result of the exchange
of $L$ between gas and stars coming from the gravity torques, the phase shift between the two 
components, and the gas dissipation \citep{bournaud02}. Depending on the amount of gas 
and the strength of the bars and spirals, the metallicity gradients can flatten in less than 1~Gyr 
(Fig.~\ref{fig:gSa}). This is in drastic contrast to the current understanding that galactic 
disks need a Hubble time for sufficient mixing (SB02, \citealt{roskar08}).  

How can we tie our results to galactic disk evolution? \cite{bournaud02} followed the 
detailed processes of bar formation, bar destruction, and bar re-formation, while 
varying the disk to bulge ratio. These authors identified three bar
formation episodes in a Hubble time. In this model, we can regard a given GalMer isolated 
galaxy simulation as one such episode of a gas accretion event during a galaxy lifetime. 
In this scenario, the rapid flattening of the metallicity gradients expected from the
vigorous migration (Fig.~\ref{fig:gSa}) would be followed by a gas enrichment at each bar 
re-formation event resulting in the rebuilding of the gradient. 
The presence or not of a metallicity gradient, or its intensity, would
then be an indicator of the bar/accretion phase of the galaxy.

The mechanism described in this study works even in low-mass galaxies (Fig.~\ref{fig:dSa}) 
and can thus for the first time provide an explanation for the extended disk profiles 
observed in galaxies with $V_c\sim100$~km/s \citep[Minchev et al. 2011, in 
preparation]{bland05}.


\acknowledgements{
Support for this work was provided by ANR-CNRS, and the AvH foundation.
}


\begin{thebibliography}{}

\bibitem[\protect\astroncite{{Bland-Hawthorn} et~al.}{2005}]{bland05}
{Bland-Hawthorn}, J., {Vlaji{\'c}}, M., {Freeman}, K.~C., and {Draine}, B.~T.:
  2005,
\newblock {\em \apj} {\bf 629}, 239

\bibitem[\protect\astroncite{{Bournaud} and {Combes}}{2002}]{bournaud02}
{Bournaud}, F. and {Combes}, F.: 2002,
\newblock {\em \aap} {\bf 392}, 83

\bibitem[\protect\astroncite{{Chilingarian} et~al.}{2010}]{chilingarian10}
{Chilingarian}, I., {Di Matteo}, P., {Combes}, F., {Melchior}, A., and
  {Semelin}, B.: 2010,
\newblock {\em \aap} {\bf 518}, A61

\bibitem[\protect\astroncite{{Dehnen}}{1998}]{dehnen98}
{Dehnen}, W.: 1998,
\newblock {\em \aj} {\bf 115}, 2384

\bibitem[\protect\astroncite{{Di Matteo} et~al.}{2007}]{dimatteo07}
{Di Matteo}, P., {Combes}, F., {Melchior}, A., and {Semelin}, B.: 2007,
\newblock {\em \aap} {\bf 468}, 61

\bibitem[\protect\astroncite{{Edvardsson} et~al.}{1993}]{edvardson93}
{Edvardsson}, B.: 1993,
\newblock {\em \aap} {\bf 275}, 101

\bibitem[\protect\astroncite{{Elmegreen} et~al.}{1992}]{elmegreen92}
{Elmegreen}, B.~G., {Elmegreen}, D.~M., and {Montenegro}, L.: 1992,
\newblock {\em \apjs} {\bf 79}, 37

\bibitem[\protect\astroncite{{Famaey} et~al.}{2005}]{famaey05}
{Famaey}, B., {Jorissen}, A., {Luri}, X., et al.: 2005,
\newblock {\em \aap} {\bf 430}, 165

\bibitem[\protect\astroncite{{Gogarten} et~al.}{2010}]{gogarten10}
{Gogarten}, S.~M.: 2010,
\newblock {\em \apj} {\bf 712}, 858

\bibitem[\protect\astroncite{{Haywood}}{2008}]{haywood08}
{Haywood}, M.: 2008,
\newblock {\em \mnras} {\bf 388}, 1175

\bibitem[\protect\astroncite{{Minchev} et~al.}{2010}]{minchev10}
{Minchev}, I., {Boily}, C., {Siebert}, A., and {Bienayme}, O.: 2010,
\newblock {\em \mnras} {\bf 407}, 2122

\bibitem[\protect\astroncite{{Minchev} \& {Famaey}}{2010}]{mf10}
{Minchev}, I. and {Famaey}, B.: 2010,
\newblock {\em \apj} {\bf 722}, 112 (MF10)

\bibitem[\protect\astroncite{{Minchev} et~al.}{2007}]{mnq07}
{Minchev}, I., {Nordhaus}, J., and {Quillen}, A.~C.: 2007,
\newblock {\em \apjl} {\bf 664}, L31

\bibitem[\protect\astroncite{{Minchev} and {Quillen}}{2006}]{mq06}
{Minchev}, I. and {Quillen}, A.~C.: 2006,
\newblock {\em \mnras} {\bf 368}, 623

\bibitem[\protect\astroncite{{Olling} \& {Dehnen}}{2003}]{od03}
{Olling}, R. and {Dehnen}, W.: 2003,
\newblock {\em \apj} {\bf 599}, 275

\bibitem[\protect\astroncite{{Quillen} and {Minchev}}{2005}]{qm05}
{Quillen}, A.~C. and {Minchev}, I.: 2005,
\newblock {\em \aj} {\bf 130}, 576

\bibitem[\protect\astroncite{{Quillen} et~al.}{2009}]{quillen09}
{Quillen}, A.~C., {Minchev}, I., {Bland-Hawthorn}, J., and {Haywood}, M.: 2009,
\newblock {\em \mnras} {\bf 397}, 1599

\bibitem[Quillen et al.(2010)]{quillen10} Quillen, A.~C., 
Dougherty, J., Bagley, M.~B., Minchev, I., 
\& Comparetta, J.\ 2010, arXiv:1010.5745

\bibitem[\protect\astroncite{{Rautiainen} and {Salo}}{1999}]{rautiainen99}
{Rautiainen}, P. and {Salo}, H.: 1999,
\newblock {\em \aap} {\bf 348}, 737

\bibitem[\protect\astroncite{{Rix} and {Rieke}}{1993}]{rix93}
{Rix}, H. and {Rieke}, M.~J.: 1993,
\newblock {\em \apj} {\bf 418}, 123

\bibitem[\protect\astroncite{{Ro{\v s}kar} et~al.}{2008}]{roskar08}
{Ro{\v s}kar}, R., {Debattista}, V.~P., {Quinn}, T.~R., {Stinson}, G.~S., and
  {Wadsley}, J.: 2008,
\newblock {\em \apjl} {\bf 684}, L79

\bibitem[\protect\astroncite{{Sch{\"o}nrich} and {Binney}}{2009}]{schonrich09}
{Sch{\"o}nrich}, R. and {Binney}, J.: 2009,
\newblock {\em \mnras} {\bf 396}, 203

\bibitem[\protect\astroncite{{Sellwood} and {Binney}}{2002}]{sellwood02}
{Sellwood}, J.~A. and {Binney}, J.~J.: 2002,
\newblock {\em \mnras} {\bf 336}, 785

\bibitem[\protect\astroncite{{Siebert} et~al.}{2010}]{siebert10}
{Siebert}, A., {Famaey}, B., {Minchev}, I., et al.: 2010,
\newblock {\em arXiv:1011.4092}

\bibitem[\protect\astroncite{{Vlaji{\'c}} et~al.}{2009}]{vlajic09}
{Vlaji{\'c}}, M., {Bland-Hawthorn}, J., and {Freeman}, K.~C.: 2009,
\newblock {\em \apj} {\bf 697}, 361

\bibitem[\protect\astroncite{{Wozniak} and {Michel-Dansac}}{2009}]{wozniak09}
{Wozniak}, H. and {Michel-Dansac}, L.: 2009,
\newblock {\em \aap} {\bf 494}, 11

\end{thebibliography}

\end{document}